\documentclass[%
reprint,
superscriptaddress,
 amsmath,amssymb,
 aps,
 pra,
]{revtex4-2}

\usepackage{color}
\usepackage{graphicx}
\usepackage{subfig}
\usepackage{dcolumn}
\usepackage{bm}
\usepackage[ruled,linesnumbered]{algorithm2e}
\usepackage{hyperref}
\usepackage{caption}
\usepackage{multirow}
\usepackage{float}

\begin{document}

\preprint{APS/123-QED}

\title{{Variational Quantum Domain Adaptation } }

\author{Chunhui WU}
\email{wch13295016367@163.com}
\affiliation{Institute of Signal Processing and Transmission, Nanjing University of Posts and Telecommunications(NUPT),
Nanjing, 210003, China.}

\author{Junhao Pei}
\affiliation{Institute of Signal Processing and Transmission, Nanjing University of Posts and Telecommunications(NUPT),
Nanjing, 210003, China.}
\author{Yihua WU}
\affiliation{Institute of Signal Processing and Transmission, Nanjing University of Posts and Telecommunications(NUPT),
Nanjing, 210003, China.}
\author{Shengmei Zhao}
\email{zhaosm@njupt.edu.cn}
\affiliation{Institute of Signal Processing and Transmission, Nanjing University of Posts and Telecommunications(NUPT),
Nanjing, 210003, China.}
\affiliation{Key Lab of Broadband Wireless Communication and Sensor Network Technology, Ministry
of Education, Nanjing, 210003, China. }

\date{\today}

\begin{abstract}
    Quantum machine learning is an important application of quantum
    computing in the era of noisy intermediate-scale quantum devices.
    Domain adaptation is an effective method for addressing the
    distribution discrepancy problem between the training data and
    the real data when the neural network model is deployed. In this
    paper, a variational quantum domain adaptation method is proposed by using
    a quantum convolutional neural network, together with a gradient reversal module,
    and two quantum fully connected layers,
     named variational quantum domain adaptation
    (VQDA). The simulations on the local computer and IBM Quantum Experience (IBM Q) platform by Qiskit show
    the effectiveness of the proposed method. The
    results demonstrate that, compared to its classical corresponding
    domain adaptation method, VQDA achieves an average
    improvement of 4 \% on the accuracy for MNIST $\rightarrow$ USPS domain transfer under
    the same parameter scales. Similarly, for SYNDigits $\rightarrow$ SVHN
    domain transfer, VQDA achieves an average improvement of 2 \% on the accuracy under
    the same parameter scales.
\begin{description}
\item[PACS numbers]
03.67.Ac,03.67.Lx

\end{description}
\end{abstract}

\maketitle


\section{Introduction}

In computer vision, the model's labeled training dataset (source domain) often has distributional differences from the real data (target domain), so the application of the model is limited.
It is also a time-consuming and laborious task to label each real data to construct a useful training set.
Therefore, domain adaptation (DA), an effective method for addressing the distribution discrepancy problem between the training data and the real data, makes the trained neural network models from the labeled source domain data effectively used in the target domain. 
Currently, there are four categories of DA methods \cite{1}, including the discrepancy-based DA, the reconstruction-based DA, the instances-based DA, and the adversarial-based DA\cite{2,3}. 
Depending on the criterion used to measure the distributional differences, the discrepancy-based DA can be further categorized into DA based on statistical criterion \cite{4}, DA based on structural criterion \cite{5},
DA based on prevalence criterion \cite{6}, and DA based on graph
criterion \cite{7}. Reconstruction-based DA is mainly implemented by
extracting domain-invariant features through a self-encoder \cite{8}.
Instances-based methods first generate the labeled target domain
samples from the labeled source domain samples, establish the
relationship between the source domain samples and the generated
samples, and then use the generated samples to train the network
suitable for the target domain \cite{9}. While the adversarial-based DA
is to introduce the game idea of generative adversarial network \cite{10}
(GAN) into the network training. For example, Ganin \emph{et al.} first proposed a
domain adversarial neural network \cite{11} (DANN) by
introducing a domain classifier and a gradient reversal layer
in the feed-forward network. When the gradient of the domain classifier
passes through the gradient reversal layer (GRL), the gradient is reversed, which enables the domain classifier
to minimize the domain classification loss at the same time as the feature extractor
maximizes the domain classification loss, and the trained network
can extract the domain-invariant features and effectively predict
the data's label.

Machine learning is one of the important applications of
quantum computing \cite{12,13}, and the resultant quantum machine learning
\cite{14,15,16,17} (QML) provides polynomial or exponential acceleration for
learning tasks. Based on the superposition and entanglement of quantum bits,
QML is expected to overcome the current problems in DA
on its large data set and slow training process \cite{18,19,20}. In the context
of the noisy intermediate-scale quantum (NISQ) era, the variational quantum
algorithm (VQA) was proposed to provide a general
framework for the implementation of QML \cite{21,22}. VQA has three steps.
The first step is to design the objective function
and the corresponding variational quantum circuit (VQC) according to
the learning task, the second step is to solve the expectation value of the objective function
by using VQC, and the third step is to
optimize the parameters in VQC by classical computation
and find the optimal parameters to satisfy the objective function \cite{23,24}.
VQA can greatly reduce the number of quantum bits, quantum gates, and the depth of required circuits by combining classical computing and quantum computing, it has become an effective way to realize quantum superiority \cite{25,26,27} nowadays.

Quantum convolutional neural network (QCNN), proposed by Cong \emph{et
al.} \cite{28}, was used to accurately identify the quantum states, and it is one
of the VQC models in VQA. Later, Hur \emph{et al.} discussed
the performance of various QCNN models in terms of the structure of
VQC, quantum data coding methods, classical data preprocessing
methods, and loss functions \cite{29}. L$\ddot{u}$\emph{ et al.} extended the application
of QCNN from quantum data to classical image data by implementing a
binary classification task in MNIST datasets \cite{30}. Yang
\emph{et al.} proposed federal learning in the context of a
decentralized feature extraction method based on QCNN to solve the
privacy preservation problem in speech recognition \cite{31}. Wei \emph{et al.} also
applied QCNN in image processing, such as image smoothing, image sharpening and edge detection \cite{32}.

Taking advantage of QCNN, we propose a QCNN-based
variational quantum domain adaptation (VQDA) method, named VQDA,
in which a new QCNN model is designed by using VQC, which has a similar hierarchical structure to a classical convolutional neural
network (CNN) model, that is, the new QCNN model has the quantum
convolutional layer, the quantum pooling layer, and the quantum fully connected (QFC)
layer. The operations on
quantum data, such as convolution, pooling, and full connection,
can be implemented by the quantum convolutional layer, the quantum
pooling layer and the quantum fully connected layer. Finally,
VQDA is achieved by introducing an additional quantum fully connected
layer for classifying domains in the
feed-forward network, and a gradient reversal module
(GRM) in the back-propagation. Taking advantage of the entanglement in the QCNN model, the proposed VQDA
outperforms the classical counterpart. We further
discuss the effectiveness of the proposed VQDA in two tasks, such as MNIST(source domain)$\rightarrow$ USPS (target domain), SYNDigits(source domain)$\rightarrow$ SVHN(target domain).

The paper is arranged as follows. In Sec.~\ref{sec:2}, we introduce the variational quantum domain adaptation and its optimization approach. In Sec.~\ref{sec:3}, we present the numerical simulations and IBM Q-platform tests of the proposed VQDA on different datasets. Finally, in Sec.~\ref{sec:4}, we draw some conclusions.

\section{\label{sec:2}Variational Quantum Domain Adaptation}

\begin{figure}[!htbp]
\centering
\includegraphics[width=1.0\linewidth]{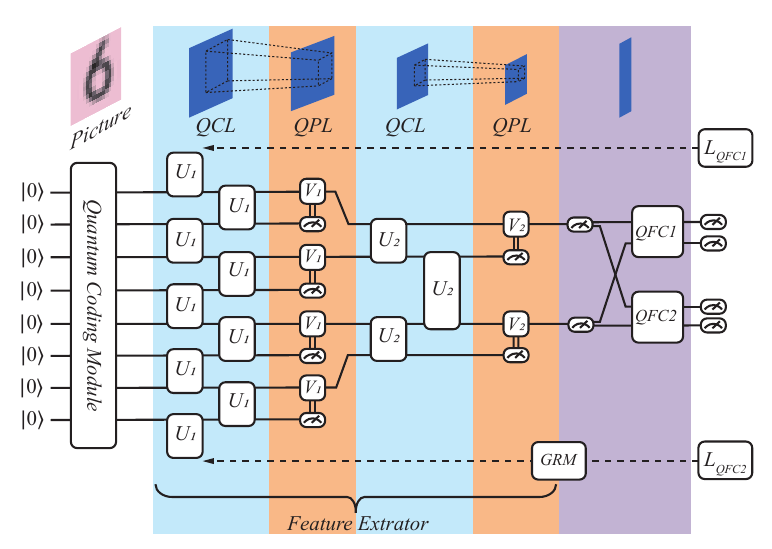}
\caption{\label{fig1} The architecture diagram of the proposed VQDA.  U is the two-quantum bit, and V is the one-quantum bit unitary transformation. QCL, QPL, QFC and GRM represent the quantum convolution layer, the quantum pooling layer, the quantum fully-connected layer and the gradient inversion module, respectively. Several alternate QCLs and QPLs comprise the `Feature Extractor'. $L_{QFC1}$, $L_{QFC2}$ are the loss functions for QFC1 and QFC2.}
\end{figure}

Inspired by the design idea of quantum convolutional neural network
\cite{28}, the quantum circuit of the proposed VQDA is shown in Fig.~\ref{fig1}.
The classical information (such as images) is firstly encoded into their corresponding quantum states through
the `Quantum Coding Module', and then the quantum convolution layer (QCL)
and the quantum pooling layer (QPL) is alternately applied to the
quantum states to extract the sample features called `Feature
Extractor', where the quantum convolution layer is comprised of several two-quantum bit unitary transformations $U_{i}$
by invariant translation, and the quantum pooling layer consists
of some measurement-control circuits that determine whether or not to
apply a single-quantum bit unitary transformation $V_{i}$ to its adjacent quantum bit.
When the system size is small enough, the remaining quantum bits
are measured to obtain the extracted features. By measuring the
quantum fully connected layer (named QFC1), the labels of the input
feature prediction samples are obtained. The `Feature Extractor'
and QFC1 construct a QCNN feed-forward network model. At the same time, an
additional quantum fully connected layer (named QFC2) is added to
VQDA for predicting the features whether from the source domain or the target domain.
QFC1 is designed to predict the labels
of the samples and QFC2 is used to predict the domains
of the samples.
In particular, the gradient of QFC2 is
multiplied by -1 through the GRM.
QFC2 and the GRM are used to ensure
that the trained `Feature Extractor' can extract the features
that cannot be distinguished from the two domains (i.e., the
common features of the two domains). The loss function
of QFC1 $L_{QFC1}$ is minimized while the loss function of QFC2 $L_{QFC2}$
is maximized for the proposed VQDA.

In the following, we describe each layer in VQDA in detail.

\subsection{Quantum Coding Module}

\begin{figure}[!htpb]
	\centering
  \includegraphics[width=1.0\linewidth]{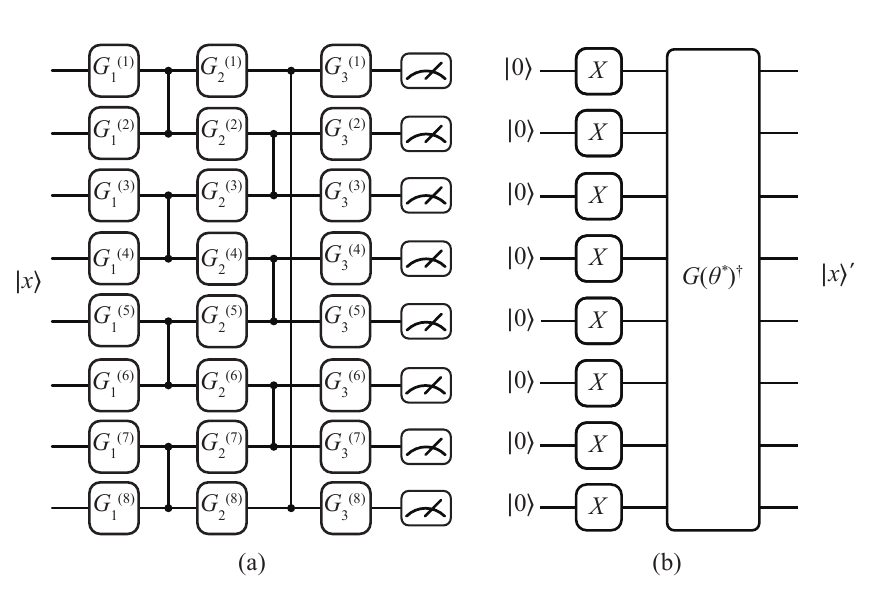}
	\caption{\label{fig2}The architect of the designed quantum circuits for `Quantum Coding Module'. (a)the parameterized alternating hierarchical quantum circuit for training, normally by numerical simulation. (b)the implementation quantum circuit. Here, $G_i^{(j)}$ represents the quantum gate, $i$ is the layer number and $j$ is the gate number.    $G\left(\boldsymbol{\theta}^\ast\right)$ denotes the trained parameter circuit in (a), $(\cdot)^\dag$ denotes the conjugate operation. X denotes the Pauli-X gate. }
\end{figure}

The first step of VQDA is preparing the quantum state of the input classical information (images). Unfortunately, there is no module to directly realize the `Quantum Coding Module' in the quantum computing cloud platform, such as the IBM Quantum Experience (IBM Q) platform. Here, we present a specific VQC to obtain the required quantum state based on the amplitude state\cite{33} preparation method proposed in \cite{34}.

The classical image can always be described by a vector $\boldsymbol {x}=\left[x_0,\cdots,x_i,\cdots,x_{2^n-1}\right]^T, i\in\{0,\cdots,2^{n}-1\}$. Its quantum state $\left|x\right\rangle$, composed by $x_i$, can then be expressed as
\begin{equation}
\left|x\right\rangle=\frac{1}{\Vert \boldsymbol {x} \Vert_2}\sum_{i=0}^{2^n-1}{x_i\left|i\right\rangle}.
\label{eq:X}
\end{equation}
where $\left|i\right\rangle$ is a computational basis, $\Vert {\boldsymbol {x}} \Vert_2$ represents the second-order norm of $\boldsymbol {x}$.
The specific VQC for the `Quantum Coding Module' is shown in Fig.~\ref{fig2}, where Fig.~\ref{fig2}(a) is the training part and Fig.~\ref{fig2}(b) is the implementation part. For the training part,  a VQC is designed to realize the transformation from $ \left|x\right\rangle $ to $\left|1\right\rangle^{\otimes n}$ by numerical simulation.
Its purpose is to pursuit the parameters ${ \boldsymbol{\theta}}$ in the VQC to perform $G\left(\boldsymbol{\theta}\right)$ on $\left|x\right\rangle$ and its resultant state is
$\left|1\right\rangle^{\otimes n}$, where $G(\cdot)$ represents the quantum operations in Fig.~\ref{fig2}(a),
$G_i^{(j)}$ in Fig.~\ref{fig2}(a) is a general single quantum bit gate composed of three quantum gates in series, that is  $G_i^{(j)}=R_Z \cdot R_Y\cdot R_Z $, $R_Z (R_Y)$ is the single quantum bit rotation gate on $Z(Y)$. All $G_i^{(j)}$ gates are presented in the alternating layered architecture, where $i$ is the layer number and $j$ is the gate number.

Although the VQC in Fig.~\ref{fig2}(a) is described as a quantum circuit, the parameters $ \boldsymbol{\theta}$ can be calculated and updated on classical computers by considering each quantum gate
operation as the matrix multiplication \cite{34}. At first, $ \boldsymbol{\theta}$ is initialized randomly,
that is,  a set of parameters $\boldsymbol{\theta}^0$ is generated by random.
Then, for a given $\left|x\right\rangle$, the set of optimal parameters $\boldsymbol {\theta}^\ast$ can be obtained when  $G\left(\boldsymbol{\theta}^\ast\right)\left|x\right\rangle=\left|1\right\rangle^{\otimes n}$ with a  gradient descent optimizer. Here, the objective function is defined as
\begin{eqnarray}
  &f\left(\boldsymbol{\theta}\right)=\frac{1}{n}\sum_{i=1}^{n}\left\langle B_i\right\rangle_{G\left(\boldsymbol{\theta}\right)\left|x\right\rangle} \\ \nonumber
  &=\frac{1}{n}\sum_{i=1}^{n}\mathrm{Tr}\left[B_iG\left(\boldsymbol{\theta}\right)\left|x\right\rangle\left\langle x\right|G\left(\boldsymbol{\theta}\right)^\dag\right],
\label{eq:f}
\end{eqnarray}
where $B_i=\boldsymbol{I}^{\otimes\left(i-1\right)}\otimes\sigma_Z\otimes \boldsymbol{I}^{\otimes\left(n-i\right)}$, $\boldsymbol{I}$ is the identity matrix, and $\sigma_Z$ is Pauli-Z matrix.
Assuming that $f\left(\boldsymbol{\theta}\right)$ can be minimized to -1, thus, all the expectation measurement values are -1, that is, the output quantum state in Fig.~\ref{fig2}(a) before the measurements can be considered as $\left|1\right\rangle^{\otimes n}$ so that the optimal parameter $\boldsymbol{\theta}^\ast$ for the unitary $G\left(\boldsymbol{\theta}^\ast\right)$ can be obtained. 

After the optimal $\boldsymbol{\theta}^\ast$ is obtained,  one can obtain the quantum state $\left|x\right\rangle$ in experimental implementation exactly by applying the circuit $G\left(\theta^\ast\right)^\dag $ on the state $\left|1\right\rangle^{\otimes n}$ based on the reversibility of quantum circuits. However, one could
not always optimize the loss $f\left(\boldsymbol{\theta}\right)$ to -1, which means the designed quantum circuits could only prepare the
amplitude encoding state approximately.
That is, only quantum state  $\left|x^{\prime}\right\rangle$ can be obtained by the quantum circuit shown in Fig.~\ref{fig2}(b), 
where $G\left(\boldsymbol {\theta}^\ast\right)^\dag$ is the conjugate quantum circuit of $G\left(\boldsymbol {\theta}\right)$, who also has the alternating layered architecture.
When the initial state $\left|0\right\rangle^{\otimes n}$ is
changed to $\left|1\right\rangle^{\otimes n}$ after $\mathrm{Pauli-}\mathrm{X}^{\otimes n}$, the quantum state $\left|x^\prime\right\rangle=G\left(\theta^\ast\right)^\dag \left|1\right\rangle^{\otimes n}$ can be achieved due to the reversibility of the quantum circuit. It is an approximation of $\left|x\right\rangle$.



\subsection{Quantum Convolutional layer}

\begin{figure}[!htpb]
  \includegraphics[width=0.9\linewidth]{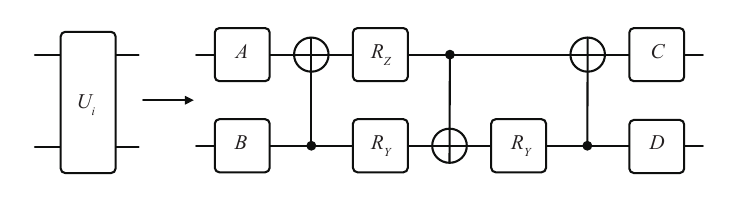}
  \caption{\label{fig3} The circuit structure of $U_i$.}
  \end{figure}

The quantum convolutional layer is designed to extract the feature of the input quantum state. 
It is comprised of some two quantum bits gate $U_i$, where $i$ denotes the convolutional layer in which it is located. Here, $U_i$ operates on the two neighboring quantum bits to demonstrate the local connectivity, as the convolutional kernel in the Convolutional neural network (CNN). At the same time,  $U_i$ is sequentially operated on each quantum bit in a translationally-invariant manner to show the parameters sharing property. To extract more features, the amplitude of the input quantum state should be flexibly adjusted by $U_i$. Thus, $U_i$ is a universal two-quantum bit circuit that can realize any unitary transformation. According to the decomposition fashion in \cite{35}, we decompose $U_i$ into a combination of a collection of basic quantum gates $\left\{\mathrm{CNOT},R_Y,R_Z\right\}$, whose circuit structure is shown in Fig.~\ref{fig3}. The expression of the proposed circuit is
\begin{equation}
  U_i=\left(A\otimes B\right)C_1^2\left(R_Z\otimes R_Y\right)C_2^1\left(I\otimes R_Y\right)C_1^2\left(C\otimes D\right),
\label{eq:Ui}
\end{equation}
where A, B, C, and D are the general single quantum bit unitary transformation, $C_j^i$ denotes a $\mathrm{CNOT}$ gate with $i$ as the control bit and $j$ as the controlled bit. As shown in Fig.~\ref{fig3}, a $U_i$ contains 15 quantum rotation gates as well as 3 $\mathrm{CNOT}$ gates.

\subsection{Quantum Pooling Layer}

In CNN, a pooling layer is usually added to the adjacent convolutional layers to reduce the feature mapping dimensions, to achieve nonlinearity, which in turn speeds up the computation, as well as prevents the overfitting. Here, the quantum pooling layer is archived by quantum measurement on some quantum bits, followed by a single quantum bit unitary transformation $V_i$ operated on each measured neighboring quantum bits, where $i$ denotes the quantum pooling layer in which it is located.

\begin{figure}[!htpb]
	\centering
  \includegraphics[width=1.0\linewidth]{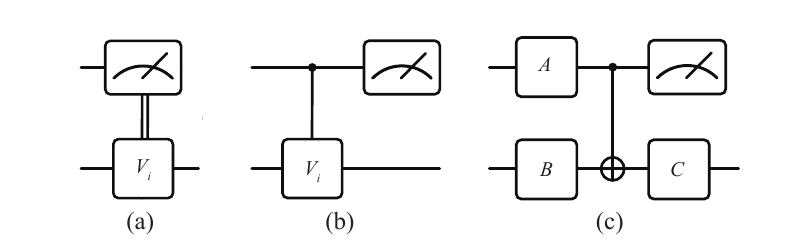}
	\caption{\label{fig4}{(a)The structure of the measurement-control circuit. (b)The equivalent circuit is based on the principle of delayed measurement. (c)The implementation of the equivalent circuit of the measurement-control circuit, where A, B and C are all the universal single quantum bit unitary transformations.}}
\end{figure}

According to the principle of deferred measurement \cite{36}, a quantum measurement can be moved to the end of a quantum circuit when it is an intermediate step in the quantum circuit and the measurement result is a condition for controlling subsequent quantum gates. 
Furthermore, to obtain the functionality of the classical pooling layer in CNN, the controlling bit should have an arbitrary control state, while the controlled bit should apply an arbitrary single quantum bit unitary transformation based on the measured results \cite{30}. Therefore, the measurement-control circuit for the quantum pooling layer can be designed as that in Fig.~\ref{fig4}. At first, the equivalent circuit for the  measurement-control circuit in Fig.~\ref{fig4}(a) can be described as that in Fig.~\ref{fig4}(b) based on the principle of the delayed measurement, then the implementation
of the equivalent circuit of the measurement-control can be realized as that in Fig.~\ref{fig4}(c).

\subsection{{Quantum Fully Connected Layer}}
\begin{figure}[!htpb]
  \includegraphics[width=0.6\linewidth]{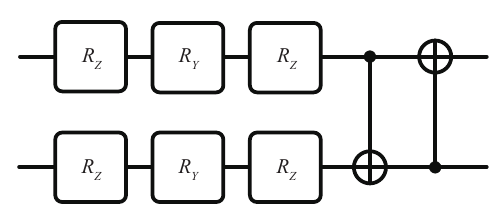}
  \caption{\label{fig5}The structure of a quantum circuit for the quantum fully connected (QFC) layer.}
\end{figure}
It is shown that the quantum circuit will have reduced quantum bits after quantum convolutional layers and quantum pooling layers. When the system size is small enough, QFC should be applied on the remained quantum bits to obtain the classification from the extracted features.
As shown in Fig.~\ref{fig1}, QFC-1 is designed to predict the label of the sample, while QFC-2 is used to predict the label of the domain.
QFC is comprised of multiple quantum circuit layers, and each quantum circuit layer consists of several quantum rotation gates and CNOT gates.
Fig.~\ref{fig5} demonstrates one quantum circuit layer for QFC, which
has 6 quantum rotation gates and 2 CNOT gates.

With the measurement of the remained quantum bits, one can obtain the expectation value, and then get the classification result after the processing on the expectation value. For example, if the classification is a binary task and the expectation values $p_1$ and $p_2$ are obtained by the quantum measurements, then the classification result $y$ can be described as

\begin{equation}
  y=\left\{\begin{matrix}0,&p_1\geq p_2\\1,&p_1<p_2\\\end{matrix}\right.
\label{eq:4}
\end{equation}
If the classification is an M-ary task, and the expectation values $p_1$,$p_2$,...,$p_M$ are obtained via quantum measurements, then the classification outcome $y$ can be described as
\begin{equation}
  y=\left\{\begin{matrix}0,p_1= max(p_1,p_2,...,p_M)\\1,p_2=max(p_1,p_2,...,p_M)\\\dots\\M-1,p_M=max(p_1,p_2,...,p_M)\\\end{matrix}\right.
\end{equation}

\subsection{Optimization in VQDA}

Since the unsupervised DA is implemented in the proposed VQDA, the labeled source domain training samples will go through QFC1 and QFC2 to get the predicted labels and the predicted domains during the forward propagation, while the unlabelled target domain training samples will only go through QFC2 to get the predicted domain. At the same time, the parameters in the quantum circuit are optimized by using the optimization method proposed in \cite{11}. That is, the gradient from QFC2 should be subtracted, instead of being added, from the feature extractor during the gradient backward propagation, so that the GRM is designed between the last quantum pooling layer and QFC2. The gradient is multiplied by -1 when the gradient from QFC2 passes through this module. Then, the optimization of VQDA can be described as
\begin{eqnarray}
&E\left(\theta_{cp},\theta_{QFC1},\theta_{QFC2}\right) =\frac{1}{n}\sum_{i=1}^{n}L_{QFC1}^{i}\left(\theta_{cp},\theta_{QFC1}\right) \\ \nonumber
    &-\frac{\lambda}{n}\sum_{i=1}^{n}L_{QFC2}^{i}\left(\theta_{cp},\theta_{QFC2}\right) \\ \nonumber
    &-\frac{\lambda}{N-n}\sum_{i=n+1}^{N}L_{QFC2}^{i}\left(\theta_{cp},\theta_{QFC2}\right).
    \label{eq:E}
\end{eqnarray}
where $n$ samples are from the source domain, $N-n$ samples are from the target domain, $L_{QFC1}^i$ denotes the $i^{th}$ sample loss on QFC1, $L_{QFC2}^i$ denotes the $i^{th}$ sample loss on QFC2, $\theta_{QFC1}$, $\theta_{QFC2}$, and $\theta_{cp}$ are the parameters in QFC1, QFC2, the quantum convolution layers and the quantum pooling layers, respectively, $\lambda$ is the domain adaptation factor.

Moreover, the rules for updating the parameters in the quantum circuit are
\begin{eqnarray}
  &\theta_{cp}\gets\theta_{cp}-\mu\left(\frac{\partial L_{QFC1}^i}{\partial\theta_{cp}}\ -\ \lambda\frac{\partial L_{QFC2}^i}{\partial\theta_{cp}}\right), \\ \nonumber
  &\theta_{QFC1}\gets\theta_{QFC1}-\mu\frac{\partial L_{QFC1}^i}{\partial\theta_{QFC1}}, \\
  &\theta_{QFC2}\gets\theta_{QFC2}-\lambda\mu\frac{\partial L_{QFC2}^i}{\partial\theta_{QFC2}}. \nonumber
\end{eqnarray}
where $\mu$ is the learning rate, and the gradient of the parameters can be obtained by the parameter shifting method \cite{37,38}. Algorithm.~\ref{algorithm:1} shows the pseudo-codes of the optimization method in VQDA.
\begin{algorithm}[!htpb]
\caption{The optimization in VQDA}\label{algorithm:1}
\SetKwData{Left}{left}\SetKwData{This}{this}\SetKwData{Up}{up}
  \SetKwFunction{Union}{Union}\SetKwFunction{FindCompress}{FindCompress}
  \SetKwInOut{Input}{input}\SetKwInOut{Output}{output}

  \Input{Source domain sample $S=\left\{\left(\left|x_i\right\rangle,y_i,d_s\right)\right\}_{i=1}^n$, target domain sample $T=\left\{\left(\left|x_j\right\rangle,d_T\right)\right\}_{j=1}^{n^\prime}$; iteration number E; batch size s; learning rate $\mu$; domain adaptation factor $\left\{\lambda\left(t\right)_{t=0}^{E-1}\right\}$; the loss function for the labelled features $L_{QFC1}$; the loss function for the domain $L_{QFC2}$}
  \Output{Optimized parameters in VQDA  $\theta^*=\left\{\theta_{cp},\theta_{QFC1},\theta_{QFC2}\right\}$}
  \BlankLine
  \emph{Randomly initialize the parameter $\theta^*$ in $\left[0,2\pi\right]$}\;
  \emph{Initialize data length $l=min\left(\left\lfloor n/s\right\rfloor,\left\lfloor n^\prime/s\right\rfloor\right)$}\;
  \For{t from 0 to E-1}{
    \For{a from 1 to l}{\label{forins}
      \emph{Randomly select s samples in S to input into the circuit}\;
      \emph{Calculate the gradient $\nabla_{\theta_{cp}^{\left(t\right)}}^SL_{QFC1}$, $\nabla_{\theta_{QFC1}^{\left(t\right)}}^SL_{QFC1}$, $\nabla_{\theta_{cp}^{\left(t \right)}}^SL_{QFC2}$ and $\nabla_{\theta_{QFC2}^{\left(t\right)}}^SL_{QFC2}$}\;
      \emph{Randomly select s samples in T to input into the circuit only through QFC2}\;
      \emph{Calculate the gradient $\nabla_{\theta_{cp}^{\left(t\right)}}^TL_{QFC2}$ and $\nabla_{\theta_{QFC2}^{\left(t\right)}}^TL_{QFC2}$}\;
      \emph{Update the parameters: $\theta_{cp}^{\left(t+1\right)}\gets\theta_{cp}^{\left(t\right)}-\mu\left(\nabla_{\theta_{cp}^{\left(t\right)}}^SL_{QFC1}-\lambda\left(t\right)\left(\nabla_{\theta_{cp}^{\left(t\right)}}^SL_{QFC2}+\nabla_{\theta_{cp}^{\left(t\right)}}^TL_{QFC2}\right)\right)$
           $\theta_{QFC1}^{\left(t+1\right)}\gets\theta_{QFC1}^{\left(t\right)}-\mu\nabla_{\theta_{QFC1}^{\left(t\right)}}^SL_{QFC1}$
           $\theta_{QFC2}^{\left(t+1\right)}\gets\theta_{QFC2}^{\left(t\right)}-\mu\lambda\left(t\right)\left(\nabla_{\theta_{QFC2}^{\left(t\right)}}^SL_{QFC2}+\nabla_{\theta_{QFC2}^{\left(t\right)}}^TL_{QFC2}\right)$}\;

    }
  }
  \emph{Output the parameters $\theta^*$}\;
\end{algorithm}

\section{\label{sec:3}simulation platforms}

In this section, we verify the performance of VQDA through numerical simulations and the experiments on the IBM Q platform for DA tasks from MNIST $\rightarrow$ USPS, from  SYNDigits $\rightarrow$ SVHN. For DA from MNIST $\rightarrow$ USPS, MNIST is the source domain which has 5000 and 1600 samples for the training set and test set, respectively; while USPS is the target domain, which has  1600 and 600 samples for the training set and test set, the samples from both domains are downsampled to a size of 16 $\times$ 16 to match 8 quantum bits VQDA model. For DA from SYNDigits $\rightarrow$ SVHN, SYNDigits is the source domain, while SVHN is the target domain. Both domains contain 1600 training and 600 test samples, and all the samples are downsampled to a size of 3-channels 16 $\times$ 16 to match the inputs of the 10 quantum bits VQDA model. The cross-entropy is chosen as the loss function for both QFC1 and QFC2, and the optimizer is Adam proposed in \cite{39}. The IBM Q quantum device is accessed through Qiskit \cite{40}. The hardware environment used for the numerical simulations is an AMD Ryzen7 4800H@2.90GHz CPU, an NVIDIA GeForce RTX 2060 (6GB) GPU, and 16G@3200MHz of RAM. All the simulation source codes are developed using Python and the Pennylane software package \cite{41}.

The number of iterations in the simulation is set to 100, and the batch size is 64, the learning rate $\mu$ is 0.001, the adopted parameter $\lambda$ for DA is given by
\begin{equation}
  \lambda=\frac{2}{1+e^{-\gamma p}}-1,
\end{equation}
where $\gamma$ = 10, $p$ is set to 0 at the beginning of training and is increased with the iterations, at last, it is approaching 1 at the end of training. The samples in MNIST, USPS, SYNDigits and SVHN are the digit numbers from 0 to 9. In the simulation, they are categorized into five classes, such as  `0' and `9', `1' and `8', `2' and `7', `3' and `6' and `4' and `5'. For the five classes, the label is 0 for the small digit, while it is 1 for the larger digit. For example, in the `3' and `6' binary classification, the label for digit `3' is 0  while it is 1 for `6'.

The proposed VQDA can be considered as the quantum version of DA discussed in \cite{11}. Therefore, we list the results with the proposed VQDA, together with that of DANN in  \cite{11} for comparison. It is noted that the parameter numbers in DANN are set to the closest to that in VQDA.

\subsection{MNIST $\rightarrow$ USPS}
\begin{figure}[!htpb]
	\centering
  \includegraphics[width=1.0\linewidth]{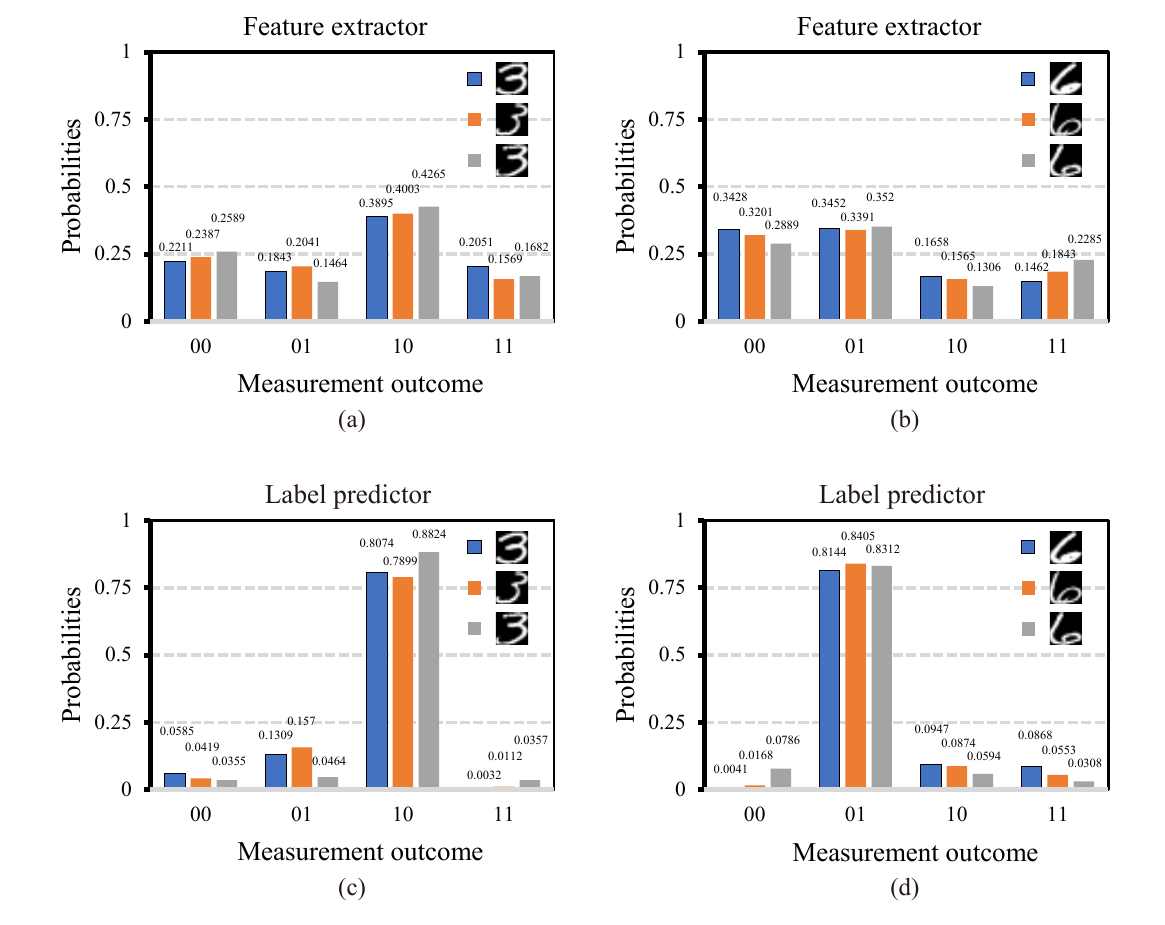}
	\caption{\label{fig6} The probability histogram after $V_2$ and QFC1 in Fig.~\ref{fig1} in USPS database on the IBM Q platform (VQDA2) with 10000 shots. (a) measurement results after $V_2$ for `3'; (b) measurement results after $V_2$ for `6'; (c) measurement results after QFC1 for `3';(d) measurement results after QFC1 for `6'.}
\end{figure}

The images in both MNIST and USPS datasets are without backgrounds.  The VQDA model used here contains two quantum convolutional layers and two quantum pooling layers, and there is a total of 246 adjustable parameters in VQDA.

At first, we demonstrate the probability histogram of $V_2$ and QFC1 in Fig.~\ref{fig1} for `3' and `6'  in the USPS database on the IBM Q platform (VQDA1) with 10000 shots in Fig.~\ref{fig6}, where the images `3' and `6' are randomly selected from the USPS database. We selected three `3' and `6' images and labeled them as `Blue', `Orange', and `Gray'. Fig.~\ref{fig6}(a) shows the measurement results after $V_2$ for the three `3', Fig.~\ref{fig6}(b) shows the measurement results after $V_2$ for the three `6', Fig.~\ref{fig6}(c) presents the measurement results after QFC1 for the three `3', and Fig.~\ref{fig6}(d) gives the measurement results after QFC1 for the three `6'.
In this case, there are two quantum measurements both at $V_2$ and QFC1, so that the measurement results are both 00,01,10 and 11.  Fig.~\ref{fig6}(a)(Fig.~\ref{fig6}(b)) shows that for different three `3'(`6'), the probability distributions after the `Feature extractor' are almost the same against the measurement outcomes. It hints that the features can be extracted by the designed `Feature extractor'. Furthermore, the results in Fig.~\ref{fig6}(c)(Fig.~\ref{fig6}(d)) demonstrate the predicted labels for the three `3'(`6') after QFC1. For the three `3', the probability of 01 outcomes is 0.8074,0.7899, and 0.8824, respectively, while the probabilities of 10 outcomes are 0.8144, 0.8405 and 0.8312, respectively, for the three `6'. It is shown that the labels for the three `3'(`6') are predicted correctly by the proposed VQDA. It is noted that the measurement outcome for the qubit is listed from right to left in Fig.~\ref{fig6}.

\begin{table}[!htpb]
  \caption{\label{tab:table1}}{The measured expectation values after $V_2$ and QFC1 in Fig.~\ref{fig1} on the IBM Q platform in USPS. }
  \begin{ruledtabular}
  \begin{tabular}{ccccccc}
   Image&$\mathrm{E}_{V_2}$-`3'&$\mathrm{E}_{QCF1}$-`3'& Result&$\mathrm{E}_{V_2}$-`6'&$\mathrm{E}_{QCF1}$-`6'&Result\\ \hline
   \multirow{2}{*}{`Blue'}&\multirow{2}{1.1cm}{0.2212 -0.1892}&\multirow{2}{1.1cm}{0.7318 -0.6212}&\multirow{2}{*}{3} &\multirow{2}{1.1cm}{0.0172 0.3760}&\multirow{2}{1.1cm}{-0.8024  0.6370} &\multirow{2}{*}{6}  \\
                         &                                   &                                   &                   &                                   &                                    &  \\\hline
   \multirow{2}{*}{`Orange'}&\multirow{2}{1.1cm}{0.2780 -0.1144}&\multirow{2}{1.1cm}{0.6636 -0.6022}&\multirow{2}{*}{3} &\multirow{2}{1.1cm}{-0.0468  0.3184}&\multirow{2}{1.1cm}{-0.7916  0.7146} &\multirow{2}{*}{6}  \\
                         &                                   &                                   &                   &                                   &                                    &  \\\hline
   \multirow{2}{*}{`Gray'}&\multirow{2}{1.1cm}{0.3708 -0.1894}&\multirow{2}{1.1cm}{0.8358 -0.8362}&\multirow{2}{*}{3} &\multirow{2}{1.1cm}{-0.1610  0.2818}&\multirow{2}{1.1cm}{-0.7240  0.8196} &\multirow{2}{*}{6}  \\
                         &                                   &                                   &                   &                                   &                                    &  \\
  \end{tabular}
  \end{ruledtabular}
\end{table}

Table~\ref{tab:table1} further lists the expectation values after $V_2$ and QFC1 in Fig.~\ref{fig6}, where
the expectation value is calculated by
\begin{equation}
\mathrm{E}= P_{\left|0\right\rangle}-P_{\left|1\right\rangle},
\end{equation}
where $P_{\left|0\right\rangle}$ represents the probability of measurement outcome $\left|0\right\rangle$, and $P_{\left|1\right\rangle}$ is the probability of measurement outcome $\left|1\right\rangle$. 
The results are obtained by Eq.(~\ref{eq:4}). It is shown that all three `3'(`6') are predicted correctly by the proposed VQDA in the `3' and `6' classifications.

Note that, for the output of `Feature Extractor', the expectation values are evaluated through Z measurement, and the obtained expectation values are then input to QFC1 in the form of angles, while the expectation values are computed through X measurement for the output of `Label Predictor'. 
When X measurement is selected, the Hadamard gate should be added before the corresponding quantum bit measurement operation. 

\begin{figure}[!htpb]
	\centering
	\centering
  \includegraphics[width=1.0\linewidth]{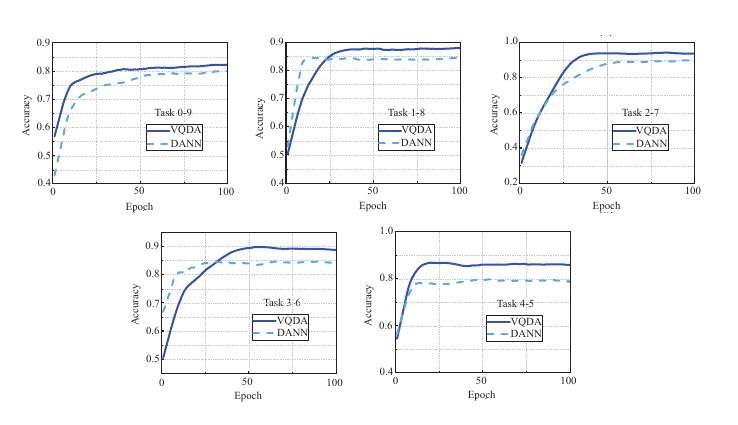}
  \caption{\label{fig7}The classification accuracy against epochs on USPS for five binary classifications.}
\end{figure}

Then, we present the classification accuracy against epochs by using VQDA (numerical simulation, VQDA2), together with those by using DANN on  USPS  during the training procedure in Fig.~\ref{fig7}. For 0-9, 1-8, and 2-7 classification tasks, VQDA2 starts to converge from the 30th iteration, while it converges at the 40th iteration for the 3-6 task and the 20th iteration for the 4-5 task. DANN get an earlier convergence than that of VQDA for 1-8 and 3-6 task and has the same as VQDA for 0-9, 2-7, and 4-5 classification tasks. Generally, the classification accuracy by VQDA2 is higher than that by DANN.

\begin{table}[!htpb]
  \caption{\label{tab:table2}}{The classification accuracy ($\%$) by VQDA and DANN for test datasets in USPS.}
  \begin{ruledtabular}
  \begin{tabular}{ccccccc}

   Model&Numbers&0-9&1-8&2-7&3-6&4-5\\ \hline
   VQDA1&246&82.70 &88.77 &94.33 &90.25 &87.24  \\
   VQDA2&246&82.67 &88.67 &94.33 &90.17 &87.17  \\
   DANN&260&80.83 &85.50 &90.33 &85.13 &81.17  \\

  \end{tabular}
  \end{ruledtabular}
\end{table}

Lastly, we discuss the classification accuracies on USPS for all the five classifications in Table~\ref{tab:table2}, together with those results by using DANN with almost the same adjustable parameters. Again, VQDA1 denotes the results in the IBM Q platform by predicting 1000 images, VQDA2 represents the numerical simulation results and
the `Number' in Table~\ref{tab:table2} denotes the parameters used in DA.
The results show that higher classification accuracy can be achieved by using VQDA than by using DANN for all five tasks. For 0-9 classification task, there is 1.87 \% accuracy improvement by VQDA1, 1.84 \% accuracy improvement by VQDA2, while there is 6.07 \% accuracy improvement by VQDA1, 6 \% by VQDA2 for the 4-5 classification task. There is an average 4 \% improvement by VQDA  over those by DANN.
It effectively demonstrates that the proposed VQDA scheme has feasibility.

\subsection{SYNDigits $\rightarrow$ SVHN}

\begin{figure}[!htpb]
	\centering
  \includegraphics[width=1.0\linewidth]{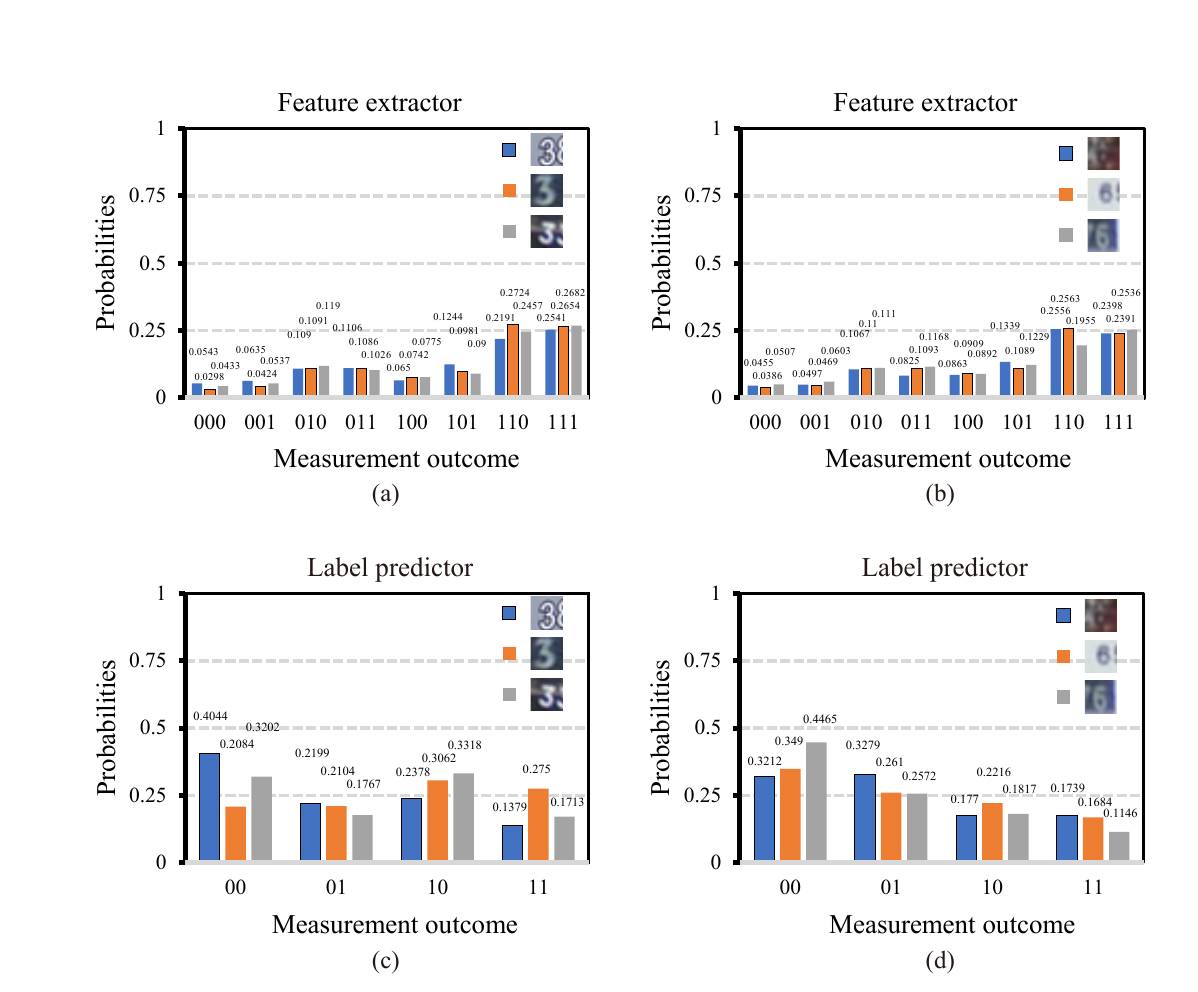}
	\caption{\label{fig8} The probability histogram of $V_2$ for `3' and `6' images in the SVHN database on the IBM Q platform (VQDA1) with 10000 shots. (a) measurement results after $V_2$ for `3'; (b) measurement results after $V_2$ for `6'; (c) measurement results after QFC1 for `3';(d) measurement results after QFC1 for `6'. }
\end{figure}
In this subsection, we testify VQDA on SYNDigits $\rightarrow$ SVHN, where SYNDigits is the source domain, and SVHN is the target domain. The images in both SYNDigits and SVHN are with backgrounds. The VQDA model uses 10 quantum bits to encode the sample information, and it contains two quantum convolutional layers and two quantum pooling layers. There are about 300 adjustable parameters inside.

Similarly, Fig.~\ref{fig8} demonstrates the probability histogram by using VQDA for 3-6 classification task in SVHN on the IBM Q platform (VQDA1) with 10000 shots, where the image `3' and `6' are with a colored background. For the 10 quantum bits circuit, the `Feature extractor' has 3 quantum measurements. The results in Fig.~\ref{fig8}(a), Fig.~\ref{fig8}(b) show that the measurement outcome probability distributions after the `Feature extractor' efficiently describe the characteristics of `3' and `6' images in SVHN. The predicted labels for the three `3'(`6') are correct by the proposed VQDA in SVHN in Fig.~\ref{fig8}(c), Fig.~\ref{fig8}(d). In the same way, the expectation values of each quantum bit after $V_2$ and QFC1 are listed in Table~\ref{tab:table3}.  It is indicated that all three colored backgrounds `3'(`6') are classified correctly by the proposed VQDA in the 3-6 classification task.
\begin{table}[!htpb]
  \caption{\label{tab:table3}}{The measured expectation values after $V_2$ and QFC1 in Fig.~\ref{fig1} on the IBM Q platform in SVHN. }
  \begin{ruledtabular}
  \begin{tabular}{ccccccc}
   Image&$\mathrm{E}_{V_2}$-`3'&$\mathrm{E}_{QCF1}$-`3'&Result&$\mathrm{E}_{V_2}$-`6'&$\mathrm{E}_{QCF1}$-`6'&Result\\ \hline
   \multirow{3}{*}{`Blue'}&\multirow{3}{1.1cm}{-0.1052 -0.3856 -0.3252}&\multirow{3}{1.1cm}{0.2844 0.2486}&\multirow{3}{*}{3} &\multirow{3}{1.1cm}{-0.0118 -0.3692 -0.4312}&\multirow{3}{1.1cm}{-0.0036  0.2982} &\multirow{3}{*}{6}  \\
                        &                                   &                                   &                   &                                   &                                    &  \\
                        &                                   &                                   &                   &                                   &                                    &  \\\hline
   \multirow{3}{*}{`Orange'}&\multirow{3}{1.1cm}{-0.0290 -0.5110 -0.4202}&\multirow{3}{1.1cm}{0.0292 -0.1624}&\multirow{3}{*}{3} &\multirow{3}{1.1cm}{-0.0084 -0.4294 -0.3904}&\multirow{3}{1.1cm}{0.1412 0.2200} &\multirow{3}{*}{6}  \\
                          &                                   &                                   &                   &                                   &                                    &  \\
                          &                                   &                                   &                   &                                   &                                    &  \\\hline
   \multirow{3}{*}{`Gray'}&\multirow{3}{1.1cm}{-0.0290 -0.4710 -0.3628}&\multirow{3}{1.1cm}{0.3040 -0.0062}&\multirow{3}{*}{3} &\multirow{3}{1.1cm}{-0.1072 -0.3538 -0.3224}&\multirow{3}{1.1cm}{0.2564 0.4074} &\multirow{3}{*}{6}  \\
                        &                                   &                                   &                   &                                   &                                    &  \\
                        &                                   &                                   &                   &                                   &                                    &  \\
  \end{tabular}
  \end{ruledtabular}
\end{table}

\begin{figure}[!htpb]
	\centering
  \includegraphics[width=1.0\linewidth]{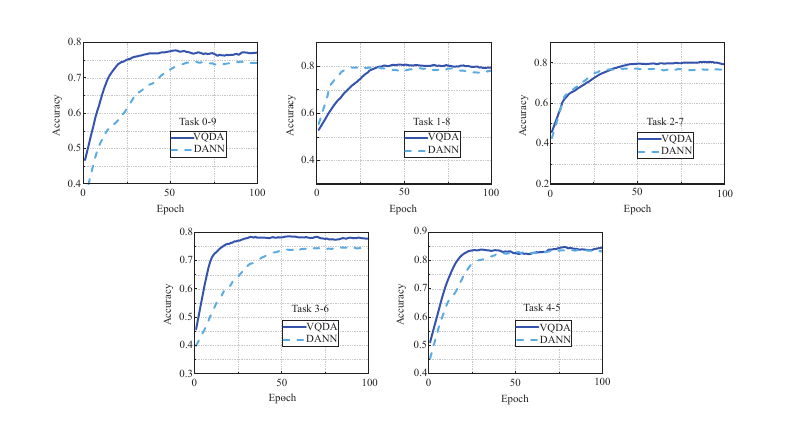}
	\caption{\label{fig9}The classification accuracy against epochs on USPS for five binary classifications in SVHN: Task 0-9; Task 1-8; Task 2-7; Task 3-6; Task 4-5.}
\end{figure}

Fig.~\ref{fig9} shows the classification accuracies of the proposed VQDA against epochs for the SVHN test set during the training process by numerical simulations (VQDA2). For 1-8 and 2-7 classification tasks, VQDA2 starts to converge from the 40th iteration, while it converges at the 20th iteration for 0-9, 3-6, and 4-5 tasks. Except for 1-8 tasks, VQDA2 has earlier convergence than that of VQDA. In general, the classification accuracy by VQDA2 is slightly higher than that by DANN.

\begin{table}[!htpb]
  \caption{\label{tab:table4}}{The classification accuracy ($\%$) by VQDA and DANN for test datasets in SVHN.}
  \begin{ruledtabular}
  \begin{tabular}{ccccccc}
   Model&Number&0-9&1-8&2-7&3-6&4-5\\ \hline
   VQDA1&300&76.85 &82.92 &79.73 &78.55 &84.65  \\
   VQDA2&300&78.94 &82.72 &81.57 &80.35 &86.15  \\
   DANN&338&76.38 &82.29 &79.79 &76.22 &85.42  \\
  \end{tabular}
  \end{ruledtabular}
\end{table}
  
Furthermore, the classification accuracies of VQDA (the IBM Q platform for  1000 images (VQDA1), the numerical simulation in database (VQDA2)) and DANN for the SVHN test set are shown in Table~\ref{tab:table4}. The parameters used in VQDA is 300, while it is 338 used in DANN. 
The results show that higher classification accuracy can be achieved by using VQDA than by using DANN for all five tasks. For the 1-8 classification task, there is 0.63 \% accuracy improvement by VQDA1, 0.43 \% accuracy improvement by VQDA2, while there is 2.33\% accuracy improvement by VQDA1, 4.13 \% by VQDA2 for the 3-6 classification task. There is an average 2 \% improvement by VQDA over those by DANN in SVHN.


\section{\label{sec:4}Conclusion}

In summary, we propose a QCNN-based DA method, named VQDA, to address the discrepancy problem between the source domain and the target domain. It comprises the quantum coding module, the quantum convolution layer, the quantum pooling layer, the quantum full-connected layer and the gradient inversion module. We discuss its availability by Qiskit on the IBM Quantum Experience (IBM Q) platform (VQDA1) and by numerical simulations on the local computer (VQDA2) for  MNIST $\rightarrow$ USPS and SYNDigits $\rightarrow$ SVHN, together with those by DANN \cite{11} with almost the same scale adjustable parameters. For MNIST $\rightarrow$ USPS,  the results by VQDA1 and VQDA2 both show that the average accuracy of VQDA is about 4 \% higher than that by DANN under the same number of parameters. For SYNDigits $\rightarrow$ SVHN,  at which time the VQDA is extended to 10 quantum bits, and the results by VQDA1 and VQDA2 show that VQDA still can achieve a higher classification accuracy when 10 quantum bits is used, there is about 2\% average accuracy improvement in comparison with those using DANN with the same scales parameter. 

\begin{acknowledgments}
  This work was supported by the National Natural
  Science Foundation of China (Grant No. 62375140).
\end{acknowledgments}

\bibliography{VQDA}

\end{document}